\makeatletter
\declare@file@substitution{revtex4-1.cls}{revtex4-2.cls}
\makeatother
\documentclass[twocolumn]{aastex631}
\shorttitle{Discriminating Babcock-Leighton-type solar dynamo models by the torsional oscillations}
\shortauthors{Congyi Zhong et al.}
\graphicspath{{./}{figures/}}
\usepackage{amsmath}
\begin{document}

%\title{Lorentz force resulting from Babcock-Leighton-type solar dynamo models and comparison with observed torsional oscillations}
\title{Discriminating between Babcock-Leighton-type solar dynamo models by torsional oscillations}

\author{Congyi Zhong}
\affiliation{ School of Space and Environment, Beihang University, Beijing, People’s Republic of China; jiejiang@buaa.edu.cn}
\affiliation{Key Laboratory of Space Environment monitoring and Information Processing of MIIT, Beijing, People’s Republic of China}

\author{Jie Jiang}
\affiliation{ School of Space and Environment, Beihang University, Beijing, People’s Republic of China; jiejiang@buaa.edu.cn}
\affiliation{Key Laboratory of Space Environment monitoring and Information Processing of MIIT, Beijing, People’s Republic of China}

\author{Zebin Zhang}
\affiliation{ School of Space and Environment, Beihang University, Beijing, People’s Republic of China; jiejiang@buaa.edu.cn}
\affiliation{Key Laboratory of Space Environment monitoring and Information Processing of MIIT, Beijing, People’s Republic of China}
\begin{abstract}
The details of the dynamo process in the Sun are an important aspect of research in solar-terrestrial physics and astrophysics. The surface part of the dynamo can be constrained by direct observations, but the subsurface part lacks direct observational constraints. The torsional oscillations, a small periodic variation of the Sun’s rotation with the solar cycle, are thought to result from the Lorentz force of the cyclic magnetic field generated by the dynamo. In this study, we aim to discriminate between three Babcock-Leighton (BL) dynamo models by comparing the zonal acceleration of the three models with the observed one. The property that the poleward and equatorward branches of the torsional oscillations originate from about $\pm 55^\circ$ latitudes with their own migration time periods serves as an effective discriminator that could constrain the configuration of the magnetic field in the convection zone. The toroidal field, comprising poleward and equatorward branches separated at about $\pm 55^\circ$ latitudes can generate the two branches of the torsional oscillations.
The alternating acceleration and deceleration bands in time is the other property of the torsional oscillations that discriminate between the dynamo models. To reproduce this property, the phase difference between the radial  ($B_{r}$) and toroidal ($B_{\phi}$) components of the magnetic field near the surface should be about $\pi/2$. 
\end{abstract}

\section{INTRODUCTION}	

The quasi 11-year cyclic variation of the Sun's large-scale magnetic fields is believed to be maintained by a dynamo process, in which the toroidal and poloidal components of the magnetic fields sustain each other via turbulent convection \citep{karak2014,Charbonneau2020}. The Babcock-Leighton (BL)-type dynamo \citep{Babcock1961, Leighton1969} is considered as the most promising one, as it has received some observational support \citep{Cameron2023}. Its poloidal field generation, i.e., the BL mechanism results from the emergence and the subsequent evolution of tilted sunspot groups over the solar surface \citep{Wang1991}. The surface part of the BL mechanism can be constrained by direct observations, and is well described by the surface flux transport model \citep{Wang1989, Baumann2004, Mackay2012, Jiang2014, Petrovay2019, Wang2020, Yeates2023}. However, the subsurface part of the dynamo process, in particular the magnetic field configuration, still lacks observational constraints.   

Helioseismology offers a unique window into the interior of our Sun and has been enormously successful in the study of the solar interior dynamics \citep[][and references therein]{Basu2016}. One of its notable successes is the identification of the latitude and depth dependence of the Sun's rotation, known as differential rotation \citep{Howe2009}. Using full-disk velocity observations over a solar cycle from the Mount Wilson observatory, \cite{Howard1980} for the first time reported the temporal variations of the rotation at different latitudes, which showed a pattern of migrating bands of faster- and slower-than-average rotation rate associated with the equatorward drift of the activity belts during the solar cycle. This pattern is known as the torsional oscillation or zonal flow. 
Since the 1990s, the development of helioseismology has gradually revealed surface torsional oscillations extend significantly downwards into the entire convection zone \citep{Kosovichev1997, Vorontsov2002, Basu2019}. There are implications that phase and amplitude of the torsional oscillations are related to magnetic activity of the subsequent solar cycle, leading some to suggest that they could serve as a proxy for the prediction of solar cycles \citep{Kosovichev2019,Ulrich2023}.

Several mechanisms have been proposed to explain torsional oscillations since their discovery. For instance, \cite{Spruit2003} proposed thermal effects near the surface to be the origin of torsional oscillations. \cite{Rempel2006} showed that the thermal effects were required to reproduce the observed equatorward branch. \cite{Beaudoin2013} demonstrated that their simulated torsional oscillations are driven via the magnetic modulation of angular-momentum transport by the large-scale meridional flow.  

However, the Lorentz force is widely considered to be the most plausible explanation for the torsional oscillations. Shortly after the discovery of the oscillations, \cite{Schuessler1981, Yoshimura1981} suggested that the torsional oscillations are driven by the Lorentz force generated by the cyclically varying magnetic field associated with the solar cycle. \cite{Yoshimura1981} estimated the Lorentz force waves using typical magnetic field values and suggested that torsional oscillations could be driven by the surface magnetic stress. Some observed properties of the torsional oscillation have been reproduced by some kinematic dynamo models. In the framework of the BL-type dynamo, \cite{Durney2000} obtained the expected sign of the longitudinal component of the Lorentz force when including the toroidal field generated in the bulk of the convection zone. \cite{Covas2000} demonstrated that torsional oscillations penetrate into the convection zone and migrate towards the equator over time. The oscillations had a period half that of the global magnetic fields. 
\cite{Chakraborty2009} attempted to explain the reason for the torsional oscillation starting 2-3 years before the solar cycle. \cite{Pipin2019} proposed that torsional oscillations can be driven by a combination of magnetic field effects acting on turbulent angular momentum transport and the large-scale Lorentz force. Their models reproduced two propagating branches in the subsurface shear layer. The effect of the Lorentz force was also investigated by a global MHD simulation of the solar interior \citep[e.g.,][]{Guerrero2016}.

There are two popular BL-type flux transport dynamo models developed by \cite{Dikpati1999} and \cite{Chatterjee2004}. Hereafter they will be referred to as DC99 and CNC04, respectively. Both models assume that the generation of the toroidal fields responsible for surface sunspot emergence occurs mainly in the tachocline. The main difference between the two models is in the flux transport parameters, especially the turbulent diffusion. The models are classified as ``advection-dominated dynamo'' and ``diffusion-dominated dynamo'' based on the relative strength between the turbulent diffusion and the meridional flow, as suggested by \cite{Yeates2008}. The different parameters result in the different structures of the magnetic field in the convection zone. \cite{Zhang2022}, hereafter referred to as ZJ22, have recently developed a BL-type dynamo that mainly operates in the bulk of the convection zone. The model takes into account some recent observational constraints, such as surface field evolution \citep{Cameron2012, Jiang2013}. At cycle minimum, the poloidal field exhibits a simple global dipole field. This generates the toroidal field with a weaker poleward and a stronger equatorward propagating branches separated around $\pm$55$^\circ$ latitudes in the convection zone.

All of the aforementioned three models can reproduce the features constrained by the surface observations, such as the butterfly diagram and the magnetic butterfly diagram. However, the interior magnetic fields' distribution and evolution of the three models differ considerably from each other. This situation raises an important question, namely, how to evaluate and discriminate between the three models. According to the original idea of \cite{Durney2000}, the torsional oscillations patterns reproduced by the magnetic field of the dynamo have the potential to discriminate between different dynamo models. This inspires us to calculate the Lorentz force and the torsional oscillations resulting from the three dynamo models, and then to compare the calculated torsional oscillations with observations so that we can constrain configurations of the magnetic field inside the convection zone by analyzing the main results from the three dynamo models.

The paper is organized as follows. Some main characteristics of torsional oscillations are described in Section \ref{sec:observations}. Three BL-type dynamo models and the relating Lorentz force are described in Section \ref{sec:model}. We present the zonal acceleration caused by the Lorentz force resulting from the three dynamo models and compare them with observations in Section \ref{sec:comparions}. We analyze how torsional oscillations are reproduced in the three models in Section \ref{sec:illustration}. We summarize and discuss our results in Section \ref{sec:conclusion}.

%%%%%%%%%%%%%%%%%%%%%%%%%%%%%%%%%%%%%%%%%%
\section{Overview of Observational Properties of the Torsional Oscillations} \label{sec:observations}
In this section, we list some main observational properties of the torsional oscillation, which will be used to discriminate among the three dynamo models.	

The generally accepted properties of the torsional oscillation are as follows. Torsional oscillations represent bands of fast and slow zonal flows that migrate towards the equator and poles, after subtracting the time-averaged rotation rate. The two migrating branches originate at about 55$^{\circ}$ latitudes. Below 55$^{\circ}$, the equatorward branch appears approximately after the previous solar activity peak and vanishes at the equator a couple of years after the minimum of the following cycle. The whole process takes about 18 years. The active regions emerge on the boundary between the fast and slow zones of the equatorward branch. Above 55$^{\circ}$, the poleward branch appears a year or so after solar minimum and moves to the pole before the next minimum. It lasts only about 9 years \citep{Howe2018,Altrock1997}, which is about half of the low-latitude branch. 

Compared to the zonal velocity, the zonal acceleration is considered to be a more representative physical quantity for revealing the physical nature of torsional oscillations \citep{Kosovichev2019,Mahajan2019}. \cite{Kosovichev2019} illustrated that the active region zones coincide with the deceleration zones. The amplitude of the zonal flow velocity is about 5 m s$^{-1}$, corresponding to the zonal acceleration on order of $10^{-8}$ m s$^{-2}$. The amplitude varies with latitudes and depth. The amplitude of the poleward branch is significantly higher than that of the equatorward branch. Torsional oscillations permeate into the solar interior and involve the whole depth of the convection zone \citep{Vorontsov2002, Mahajan2019}. The amplitude of torsional oscillations decreases with increasing depth.

Some properties of the torsional oscillation are still under debate. For example, the phase shift of the low-latitude branch with depth remains uncertain. \cite{Kosovichev2019} point out that at low latitudes, acceleration bands first appear at the bottom of the convection zone and rise to the near surface for about 8-9 yrs at about 15$^{\circ}$ latitude. At higher latitudes, there is a similar phase shift process, but the rate is much faster. It is about 2 yr at 60$^{\circ}$. However, \cite{Basu2019} present an opposite trend for 15$^\circ$ and 30$^\circ$ latitudes. That is, acceleration bands first appear near surface and propagate downward with time (see the top two panels of their Figure 4). \cite{Howe2005} suggest that the phase of the torsional oscillations has no obvious regularity at low latitudes, but it is almost unchanged at high latitudes.

\section{BL-type Dynamo Models and the Associated Lorentz Force} \label{sec:model}
\subsection{BL-type Dynamo Models} \label{sec:BL-model}

All the three BL-type dynamo models, i.e., DC99, CNC04, and ZJ22, are kinematic dynamo models. The starting point of the models is the magnetic induction equation, governing the evolution of the large-scale magnetic field $\textbf{\textit{B}}$ in response to advection by the flow field $\textbf{\textit{U}}$ and magnetic diffusion $\eta$,
\begin{equation}
	\frac{\partial \textbf{\textit{B}}}{\partial t}=\nabla\times(\textbf{\textit{U}}\times\textbf{\textit{B}}-\eta\nabla\times\textbf{\textit{B}}).\label{eq:inductionEq}
\end{equation}	
Under the assumption of axisymmetry, the magnetic field can be written as $\textbf{\textit{B}}(r,\theta,t)=B_{\phi}(r,\theta,t) \hat{\textbf{\textit{e}}}_\phi+
\nabla\times
\left[A(r,\theta,t)\hat{\textbf{\textit{e}}}_\phi\right]$, where $B_{\phi} \hat{\textbf{\textit{e}}}_{\phi}$ represents the toroidal field and $\textbf{\textit{B}}_p = \nabla\times\left[A(r,\theta,t)\hat{\textbf{\textit{e}}}_\phi\right]$ is the poloidal component. The dynamo equations resulting from the induction equation are expressed as, 
\begin{equation}
	\frac{\partial A}{\partial t}+\frac{1}{s}(\textbf{\textit{u}}_{p}\cdot\nabla)(sA)
	=\eta\left(\nabla^{2}-\frac{1}{s^{2}}\right)A+S_{BL},\label{eq:eq1}
\end{equation}
\begin{equation}
\begin{split}
    &\frac{\partial B_{\phi}}{\partial t}+\frac{1}{r}\left[\frac{\partial(u_{r}rB_{\phi})}
	{\partial r}+\frac{\partial(u_{\theta}B_{\phi})}{\partial\theta}\right]\\
        &=\eta\left(\nabla^{2}-\frac{1}
	{s^2}\right)B_{\phi}+s(\textbf{\textit{B}}_{p}\cdot\nabla\Omega)+\frac{1}{r}\frac{d\eta}
	{dr}\frac{\partial(rB_{\phi})}{\partial r},\label{eq:eq2}
\end{split}
\end{equation} 
where $S_{BL}$ is the BL-type source term and $s=r\sin \theta$. The flow fields $\textbf{\textit{U}}$ include azimuthal rotation $\textbf{\textit{U}}_\phi=\Omega r\sin\theta\hat{\textbf{\textit{e}}}_{\phi}$ and meridional flow $\textbf{\textit{u}}_{p} = u_r\hat{\textbf{\textit{e}}}_{r}+u_{\theta}\hat{\textbf{\textit{e}}}_{\theta}$. The three models are distinguished by specific assumptions for the key ingredients of the dynamo process, such as the magnetic diffusion $\eta$, the source term of the poloidal fields, and the flow fields. Below we will overview the main differences among the three models.

In the dynamo model, the $\Omega$-effect describes the process that the poloidal fields are sheared by differential rotation to generate the toroidal fields. In the models of DC99 and CNC04, they all assume that the $\Omega$-effect mainly works in the tachocline, which contains strong radial shear and subadiabatic stratification. Since this scenario faces challenges of MHD simulations and stellar observations \citep{Nelson2014,Wright2016}, ZJ22 developed a model, in which the toroidal fields are mainly generated in the bulk of the convection zone by the latitudinal differential rotation. The tachocline plays a negligible role in the ZJ22 model. 

For the BL-type source term, the $\alpha$-effect, the DC99 model adopts a non-local surface source term that relates to the toroidal fields at the bottom of the convection zone. The CNC04 model adopts a buoyancy algorithm that makes the toroidal fields at the bottom of the convection zone rise to the surface when these toroidal fields are larger than a threshold, then the surface BL source term works to generate the newly poloidal fields. The ZJ22 model makes the surface BL-type source term related to the subsurface toroidal flux through the whole bulk of the convection zone.

For the profiles of differential rotation used in the three models, they are almost the same since they all adopted the converged helioseismic results. However, the inner profile of the meridional flow has not reached a consensus so far. So the profiles of the meridional flow used in these models are different. The meridional flow in the DC99 model penetrates into about 0.6 $R_\odot$. It is about 0.61 $R_\odot$ in CNC04. These mean that the meridional flow penetrates below the bottom of the convection zone in both models. In ZJ22 the penetration depth is about 0.7 $R_\odot$. The amplitudes of the meridional flow at the surface are 10, 29, and 20 m s$^{-1}$ in DC99, CNC04, and ZJ22, respectively. 

In the DC99 and CNC04 models, the meridional flow plays an important role in coupling the regions of the $\alpha$ and $\Omega$-effects and in determining the cycle period. In the ZJ22 model, however, the role of the meridional flow is much weaker \citep{Zhang2024}. The surface poloidal fields is in radial, and its evolution is consistent with observations. In the convection zone, the poloidal fields have the regular large-scale structure, approximately to the dipolar structure. The latitudinal propagation of the toroidal flux belts \citep{Zhang2022b} is dominated by the latitude dependence of the latitudinal rotational shear, which peaks around $\pm$55$^\circ$ latitudes. This leads to two branches of the toroidal field: a weaker poleward branch and a stronger equatorward branch influenced by the equatorward component of the meridional flow.

The turbulent diffusivity is another key ingredient in the dynamo models. According to the mixing length theory, the diffusivity can reach $10^{13}$ cm$^2$ s$^{-1}$ in the bulk of the convection zone. However, usually dynamo models cannot work with such high diffusivity. In the reference case of DC99, the diffusivity is  5$\times10^{10}$ cm$^2$ s$^{-1}$. The reference case of CNC04 operates in the relatively high diffusivity range of $10^{12}$ cm$^2$ s$^{-1}$. But the strong diffusivity is just for the poloidal field. The diffusivity for the toroidal fields is still two orders lower than that of the poloidal fields. The diffusivity for both the toroidal and the poloidal fields is 2$\times10^{11}$ cm$^2$ s$^{-1}$ in the model of ZJ22. The strength of diffusion in the convection zone impacts the configuration of magnetic fields, which further determine various properties of the Lorentz force.

\subsection{The Lorentz Force and Zonal Acceleration} \label{sec:LorentzForce}
In the kinematic dynamo presented in the last subsection, the effect of the magnetic field on the velocity is ignored and all the flow fields are time-independent. In this case, the steady profile of velocity field is determined by the Navier–Stokes equation in equilibrium. Since we only investigate the time variation of the rotation, only the azimuthal direction, i.e., the $\phi$ direction in the spherical coordinate ($r$, $\theta$, $\phi$) satisfies:
%\textbf{The evolution of the toroidal velocity field \textbf{\textit{U}} is given by the Navier–Stokes equation:}
%The evolution of the velocity field \textbf{\textit{U}} in the presence of the magnetic field \textbf{\textit{B}} is given by the Navier–Stokes equation: 
\begin{equation}
	\rho \frac{\partial U_\phi}{\partial t} = -\left[(\textbf{\textit{U}}\cdot\nabla)\textbf{\textit{U}}\right]_\phi-(\nabla p)_\phi+(\nu\nabla^2\textbf{\textit{U}})_\phi = 0, \label{eq:NS0}
\end{equation}
where $\rho$ is density, $p$ is pressure, and $\nu$ is the turbulent viscosity.
When the effect of the magnetic field $\textbf{\textit{B}}$ on the velocity, that is the Lorentz force, is taken into account, there is not only the steady profile of the toroidal velocity field, but also a perturbed part $V$ denoting to the torsional oscillation. The temporal variation of this part satisfies
%When the magnetic field $B$ is taken into account, a perturbed part $V$ of the toroidal velocity field denoting to the torsional oscillation is added to the steady profile of the velocity field. The temporal variation of this part is

\begin{equation}
	\rho \frac{\partial V}{\partial t} = -\delta\left[(\textbf{\textit{U}}\cdot\nabla)\textbf{\textit{U}}\right]_\phi-\delta(\nabla p)_\phi+\delta(\nu\nabla^2\textbf{\textit{U}})_\phi+F_{\phi}, \label{eq:NS1}
\end{equation}
where $\delta$ describes the difference between the cases with and without the magnetic field and $F_\phi$ is the Lorentz force in the $\phi$ direction. The difference of momentum transfer term $\delta\left[(\textbf{\textit{U}}\cdot\nabla)\textbf{\textit{U}}\right]_\phi$ is small enough in comparison to the Lorentz force so that it can be ignored, which follows the assumption given by \cite{Yoshimura1981}.
We further assume axisymmetry, that is $\partial/\partial \phi=0$. Hence the terms involving the zonal gradient are zero. Then Equation (\ref{eq:NS1}) is reduced to

\begin{equation}
	\rho\frac{\partial V}{\partial t}	=F_\phi .\label{eq:F}
\end{equation}

The toroidal Lorentz force $F_{\phi}$ is expressed as
\begin{equation}
    F_{\phi} = \frac{1}{4\pi}\left[(\nabla\times \textbf{\textit{B}})\times\textbf{\textit{B}}\right]_{\phi}, \label{eq:LorenzForce0}
\end{equation}
which can be rewritten with the three components of the magnetic field as
\begin{equation}
    F_{\phi}=\frac{1}{4\pi}\left[  B_r\frac{\partial B_{\phi}}{\partial r}+\frac{B_{\theta}}{r}\frac{\partial B_{\phi}}{\partial \theta}+\frac{B_{r}B_{\phi}}{r}+\frac{\cot\theta B_{\theta}B_{\phi}}{r}\right]. \label{eq:LorenzForce1}
\end{equation}

Equations (\ref{eq:LorenzForce0}) and (\ref{eq:LorenzForce1}) indicate that not only the strength but also the configuration of the magnetic fields impact the value of $F_{\phi}$. This explains why torsional oscillations occur in regions without strong magnetic fields \citep{Howard1983}.

According to Equation (\ref{eq:F}), the zonal acceleration $\partial V/\partial t$ caused by the Lorentz force can be calculated using the given density profile. We adopt the density stratification based on a least-squares fit to the solar model S as \citep{Christensen1996}
\begin{equation}
    \rho(r) =0.5* (R_{\odot}/r-\gamma)^{m}, \label{eq:rho}
\end{equation}
where $\gamma = 0.9665$ and $m$ = 1.911 based on the constraint by \cite{Jaramillo2009}. The density near the surface is much smaller than that at the bottom of the convection zone. Therefore, even though the magnetic fields in the shallow convection zone are smaller than that at bottom, it still can support the same amplitude of the zonal acceleration.

\section{Results} \label{sec:results}
%\subsection{Zonal acceleration caused by the Lorentz force and comparison with observations}
\subsection{Torsional Oscillations from the Three Models and Comparison with Observations}
\label{sec:comparions}

\begin{figure*}[ht!]
	\centering
	\includegraphics[width=18 cm]{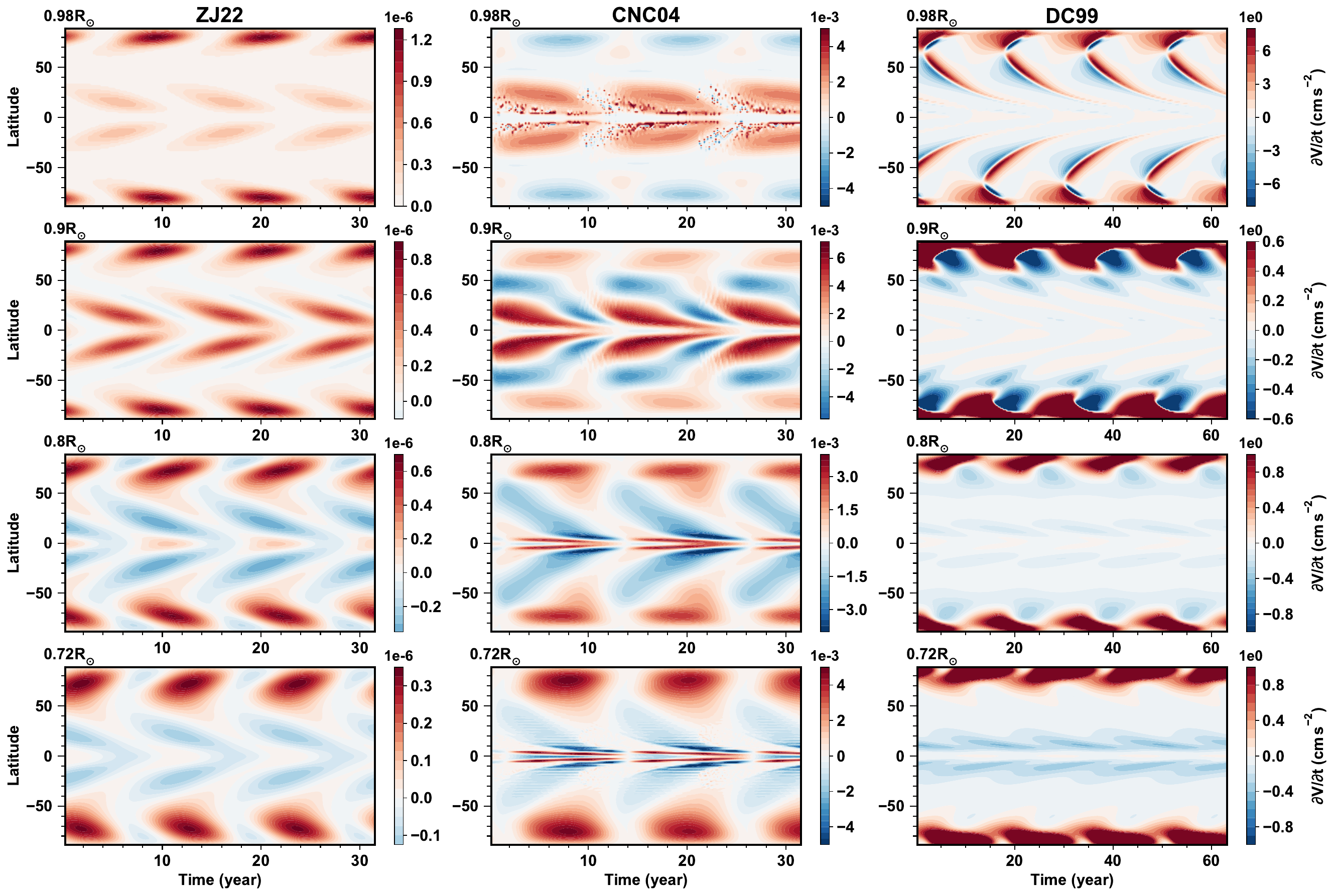}
	\caption{Time-latitude diagrams of the zonal acceleration, $\partial V/\partial t$, driven by the Lorenz force for ZJ22 (left column), CNC04 (middle column), and DC99 (right column) at four different depths. Four rows from top to bottom correspond to the results in depth of 0.98R$_\odot$, 0.9R$_\odot$, 0.8R$_\odot$, and 0.72R$_\odot$, respectively.\label{fig1}}
\end{figure*}

In this subsection, we will show torsional oscillations calculated based on the three dynamo models, and then compare the results with observations in the three aspects: 1) equatorward and poleward migrating branches including the propagation time period, 2) alternating acceleration and deceleration bands in time, and 3) amplitude. Figure \ref{fig1} shows the time-latitude diagrams of the zonal acceleration driven by the Lorenz force in the models of ZJ22, CNC04, and DC99, in the depth of 0.72$R_\odot$, 0.8$R_\odot$, 0.9$R_\odot$, and 0.98$R_\odot$.

%\iffalse 	
\unskip
%\fi

Observations of torsional oscillations show distinct equatorward and poleward branches as presented in Section \ref{sec:observations}. This provides the first property to discriminate the three models. 
In the ZJ22 model, the poleward and equatorward branches separate around $\pm 55^{\circ}$, which corresponds to the initial latitude of torsional oscillations. In the bulk of the convection zone, the equatorward branch exists for about 16 years, and the poleward branch lasts only about 8 years. The propagation time period of each torsional oscillation branch generated by the ZJ22 model is in close agreement with the observed results.
The CNC04 model is able to generate a torsional oscillation pattern that exhibits both propagating branches near the surface. The equatorward branch lasts about 12 years, which is shorter than the observations, and poleward branches persist for about 10 years. However, the equatorward branch occurs at high latitudes in the bulk of the convection zone and the poleward branch is absent.
In the DC99 model, the poleward branch above $\pm 70^{\circ}$ latitudes last about 15 years throughout the convection zone. Migration patterns of the equatorward branch are more complex. The surface equatorward branch propagates from about $\pm 70^{\circ}$ latitudes with time scale of 30 years. In the bulk of the convection zone, the equatorward branch generates in two regions. One is slightly below the poleward branch and the other is below $\pm 20^{\circ}$ latitudes. It is worth noting that since the solar cycle period of the DC99 model we reproduced is about twice that given in \cite{Dikpati1999}, we increased the time frame in Figure \ref{fig1} accordingly.

The second property to discriminate the dynamo models is the alternating acceleration and deceleration bands. In the ZJ22 model, at $0.9R_{\odot}$ and above, the two propagating branches are dominated by acceleration zones. Below $0.9R_{\odot}$, the deceleration band at low latitudes and acceleration band at high latitudes are visible. In the CNC04 model, the surface equatorward branch is dominated by acceleration bands while the poleward branch has both acceleration and deceleration zones. As the depth increases, the deceleration band gradually dominates the equatorward branch. The surface migrating branches has alternating bands in the DC99 model. In the subsurface, the poleward branch is dominated by acceleration bands, while the equatorward branch is governed by deceleration bands.

Another property of torsional oscillation is its amplitude. \cite{Kosovichev2019} shows the acceleration of torsional oscillation is in order of $10^{-6}$ cm s$^{-2}$.
%\textbf{\cite{Durney2000} gives the estimation of the amplitude of the zonal acceleration, which is in order of $10^{-6}$ cm s$^{-2}$.} 
The results from the ZJ22 model match the observed estimation in order of magnitude. In contrast, the results from the models of CNC04 and DC99 are 3 orders and 6 orders stronger than observations, respectively. We note that the reason that \cite{Chakraborty2009} show a good consistency with the observed one is that a filling factor $f$=0.067 was introduced in their calculations. The ZJ22 model also shows that the poleward branch has higher amplitude than the equatorward branch, which is also a well-reproduced property of the observed torsional oscillation. 

The amplitude of the torsional oscillation provides the third property to discriminate between the dynamo models. The amplitude produced from the ZJ22 model matches the observed amplitude of the torsional oscillations. The polar field near the surface is about few Gauss, which is consistent with the observation, and the toroidal field in the bulk of the convection zone in the ZJ22 model is several hundred Gauss. This might provide a clue about the strength of toroidal field in the solar convection zone. The CNC04 and DC99 models produce the toroidal field in the order of $10^4-10^5$ G. This high toroidal field strength leads to the high amplitudes of the acceleration bands. The strength of toroidal field in order of hundred Gauss is sufficient to reproduce the observed amplitude of torsional oscillations. The toroidal magnetic flux corresponding to the toroidal field of this strength can be up to the order of $10^{23}$ Mx in a hemisphere, which is in order of magnitude agreement with the constraint given by \cite{Cameron2015}.

%\newpage
\subsection{Understanding of the Main Results of Torsional Oscillations from the Three Models}
\label{sec:illustration}
In the last subsection, we have presented that the torsional oscillations from the three dynamo models and compared them with the three observed properties of torsional Oscillations, namely, equatorward and poleward migrating branches including the propagation time period, alternating acceleration and deceleration bands in time, and amplitude. In this subsection we will analyze how the first two properties are formed and will further provide constraints on the solar interior magnetic field based on the comparisons.

\subsubsection{Alternating Acceleration and Deceleration Bands in Time}
Alternating acceleration and deceleration bands in time is one of the prominent properties of the torsional oscillations occurring throughout the convection zone. Here we put emphasis on understanding the property near surface.

%\iffalse 
\begin{figure}[htp]
%        \centering
	\includegraphics[width=9 cm]{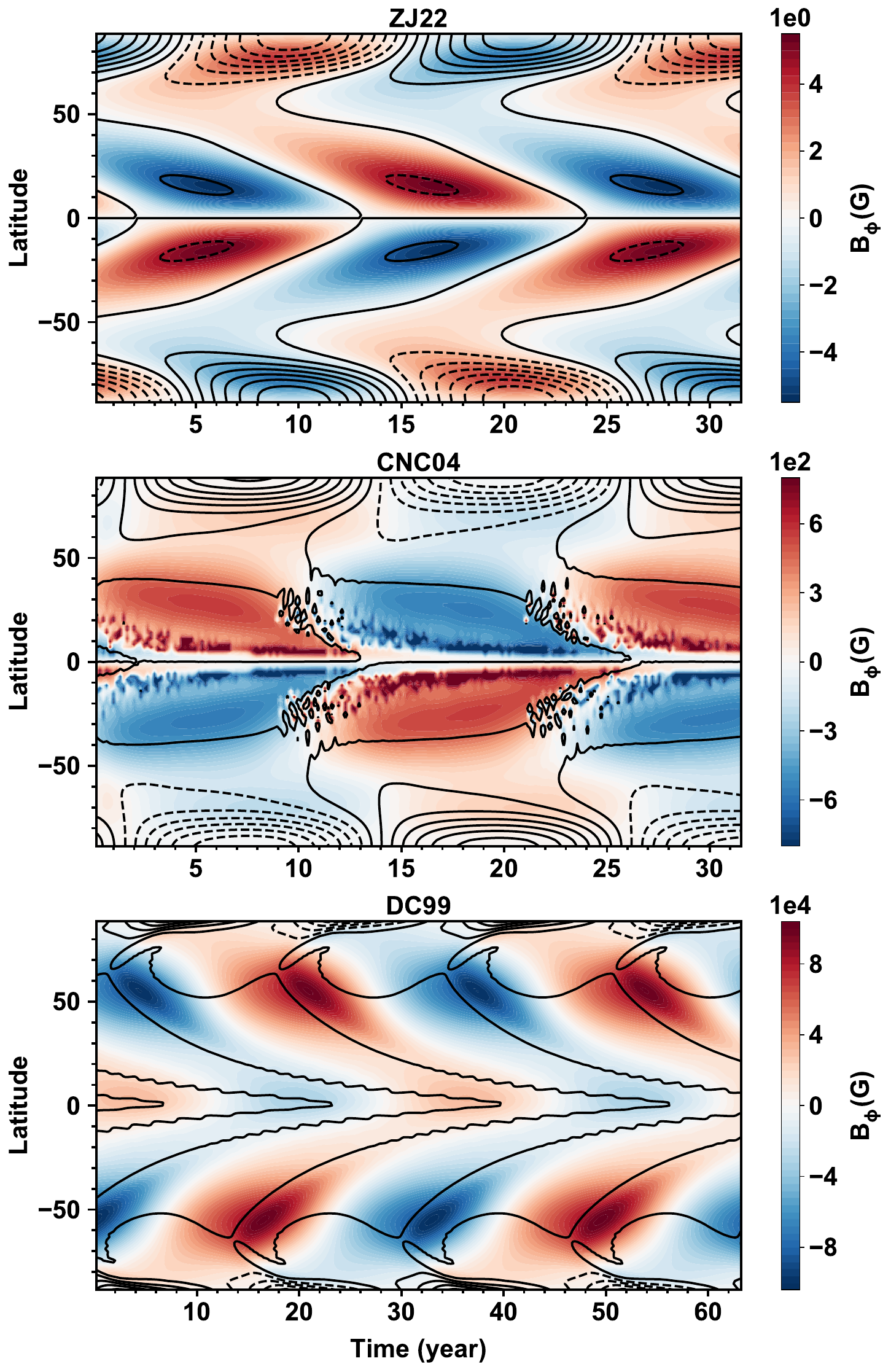}
	\caption{ Time-latitude diagrams of the surface toroidal field (color shades) and radial field (contours) for ZJ22 (top panel), CNC04 (middle panel), and DC99 (bottom panel). The solid and dashed lines of the contours denote positive and negative values.
		\label{fig3}}
\end{figure}   
\unskip
%\fi

According to Equation (\ref{eq:LorenzForce1}), the last three terms of the zonal Lorentz force $F_{\phi}$ vanish at the solar surface since $B_{\phi}$ is zero determined by the outer boundary conditions of dynamo models. Therefore, we have
\begin{equation}
	F_{\phi}\bigg|_{r \to R_{\odot}} \approx\frac{1}{4\pi} B_r\frac{\partial B_{\phi}}{\partial r} \approx -\frac{1}{4\pi} \frac{B_r B_{\phi}}{R_{\odot}-r}. \label{eq:simplifyLF}
\end{equation}	
Here $r$ is chosen to be near the solar surface $R_{\odot}$ in order to ensure the validity of the simplification. The zonal Lorentz force $F_{\phi}$ exhibits an opposite sign compared to the product of $B_{r}$ and $B_{\phi}$. When $B_{r}B_{\phi}>0$, $F_{\phi}$ is in the opposite direction of the differential rotation, resulting in a deceleration band for the torsional oscillations. Conversely, when $B_{r}B_{\phi}<0$, the effect is reversed. Thus, the torsional oscillations provide a constraint on the phase relationship between the two components of the magnetic field from dynamo models. 

To understand the alternating acceleration and deceleration bands near the surface, we give the time-latitude diagram of the toroidal field $B_{\phi}$ superposed by the contour of radial field $B_{r}$ at $0.98R_{\odot}$ in Figure \ref{fig3}. Thus, the phase relation between $B_{r}$ and $B_{\phi}$ can be illustrated.
Figure \ref{fig3}(a) corresponds to the result of the ZJ22 model. The phase difference between $B_{r}$ and $B_{\phi}$ is about $\pi$. This results in $B_{r}B_{\phi}<0$ and $F_{\phi}$ in the direction of the rotation  persistently. This explains the absence of the deceleration band of the near-surface torsional oscillations from the ZJ22 model.
The phase difference between $B_{r}$ and $B_{\phi}$ in the CNC04 model (middle panel) is about $\pi$ at low latitudes, while it is less than $\pi/2$ and lager than $0$ at high latitudes. This leads to the absence of deceleration zone of equatorward branches and the narrow acceleration of poleward branches. In the DC99 model (right panel), the phase difference of approximately $\pi/2$ exists between $B_{r}$ and $B_{\phi}$. This indicates that the sign of $B_{r}B_{\phi}$ alternates between positive and negative values, consequently leading to the alternating positive and negative zonal Lorentz force.

In summary, to achieve torsional oscillations characterized by alternating acceleration and deceleration bands near the solar surface, the phase difference between $B_{r}$ and $B_{\phi}$ should be approximately $\pi/2$. This also corresponds to that $B_{r}B_{\phi}$ changes its sign periodically near the surface. Note that \cite{Stix1976} once suggested the relation $B_{r}B_{\phi}<0$ holds at the activity belt near the surface based on the observed magnetic butterfly diagrams. However, as pointed out by \cite{Schussler2005}, the phase relation given by \cite{Stix1976} cannot be taken as a constraint for solar dynamo models. The emergence of tilted active regions and flux transport through surface flows \citep{Jiang2014, Yeates2023} naturally and inevitably leads to the phase relation. Our constraint on the phase relation between the poloidal and the toroidal magnetic field components has totally different origins from that given by \cite{Stix1976}.

In the deep convection zone, other terms of Equation (\ref{eq:LorenzForce1}) could play a dominant role in determining  $F_{\phi}$, which will be presented in the next subsection. It is inappropriate to adopt the simplified form presented in Equation (\ref{eq:simplifyLF}) anymore. Note that \cite{Chakraborty2009} only considered the first term of Equation (\ref{eq:LorenzForce1}) to calculate and analyze $F_{\phi}$ in the convection zone based on the CNC04 model. 

\subsubsection{Poleward and Equatorward Branches Originating from about $\pm 55^\circ$ Latitudes}
\label{sec:twoBranches}
The other prominent property of the torsional oscillations is that they have poleward and equatorward branches originating from about $\pm 55^\circ$ latitudes with different migration time periods. Only results from the ZJ22 model are consistent with the property overall. Here we take the depth of 0.8$R_\odot$ as an example to analyze the reasons.

\begin{figure}[htp]
%        \centering
	\includegraphics[width=9 cm]{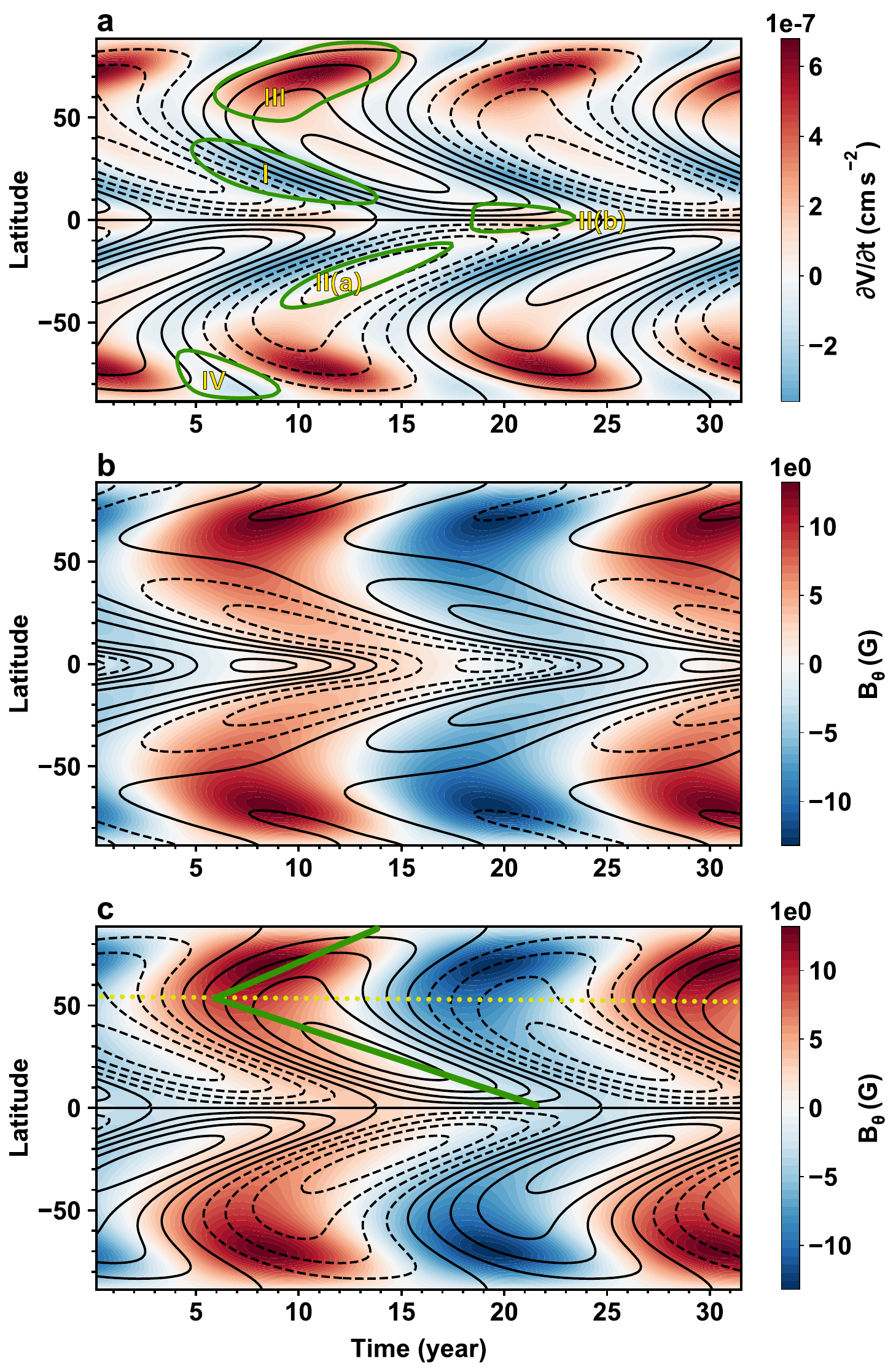}
	\caption{ Time-latitude diagrams of the torsional oscillation and magnetic field information for the ZJ22 model at $0.8R_{\odot}$. (a) The torsional oscillation (color shades) and $B_{\phi}$ (contours). The green circles correspond to the periodic migration bands. (b) $B_{\theta}$ (color shades) and $\partial B_{\phi}/\partial \theta$ (contours). (c) $B_{\theta}$ (color shades) and $B_{\phi}$ (contours). The green polyline, referred to as the `$\textless$' type in the main text, corresponds to the propagating branches of $B_{\phi}$. The yellow dotted lines correspond to the $55^\circ$ latitude. The solid and dashed lines of the contours denote positive and negative values.}
		\label{fig:ExplTwoBranches_ZJ22}
\end{figure}

Figure \ref{fig:ExplTwoBranches_ZJ22}(a) shows the time-latitude diagram of the torsional oscillations at 0.8$R_\odot$ and the corresponding toroidal field from the ZJ22 model. The $55^\circ$ latitude and the periodic migration pattern in two directions divide the diagram into four bands as presented by Figure 6(d) of  \cite{McIntosh2014} about the SOHO/MDI differential rotation residual. Note that the parameters are either symmetric or anti-symmetric about the equator. To avoid overcrowding, we mark the four bands separately in two hemispheres. 

%\iffalse 
\unskip

\begin{figure*}[!htp]
	\centering
	\includegraphics[width=18 cm]{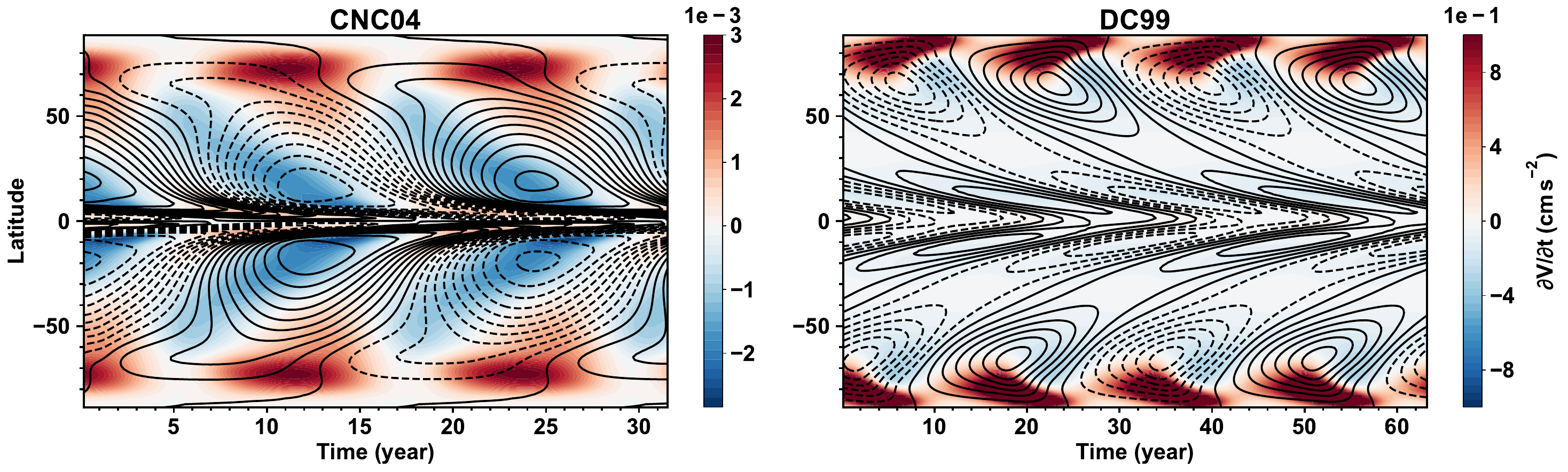}
	\caption{Time-latitude diagrams of the torsional oscillation (color shades) and the toroidal field (contours) at $0.8R_{\odot}$ for CNC04 (left panel) and DC99 (right panel). The solid and dashed lines of the contours denote positive and negative values.
		\label{fig:ExplTO_CNC_DC}}
\end{figure*}  
\unskip

Band (I) corresponds to the equatorward deceleration band of the torsional oscillations with the cycle period of about 16 years. It exactly follows the equatorward migration of the toroidal field from about the $55^\circ$ latitude to the equator. The band is distributed along the boundary of each two consecutive cycles. And the equatorward deceleration occurs when and where $\partial B_{\phi}/\partial \theta$ is large and $B_{\phi}$ is small. Moreover, the ZJ22 model's poloidal field is dominated by the latitudinal component $B_{\theta}$, and $B_r$ is nearly zero in the non-polar regions \citep{Jiang2013}. Hence the second term of Equation (\ref{eq:LorenzForce1}) dominates $F_{\phi}$, which is 
\begin{equation}
	F_{\phi}\approx\frac{1}{4\pi}\frac{B_{\theta}}{r}\frac{\partial B_{\phi}}{\partial \theta} ,
	\label{eq:force_2nd_term}
\end{equation}
and is opposite to the rotation direction. Figures \ref{fig:ExplTwoBranches_ZJ22}(b) and (c) are the time-latitude diagrams of $B_{\theta}$ overlapped with the contours of $\partial B_{\phi}/\partial \theta$ and $B_{\phi}$ at $0.8R_{\odot}$, respectively. Figure \ref{fig:ExplTwoBranches_ZJ22}(b) shows that when $B_{\theta}$ changes its sign about every 11 years, $\partial B_{\phi}/\partial \theta$ changes sign simultaneously. Hence the equatorward deceleration band occurs during each transition of two consecutive cycles. In the ZJ22 model, the latitudinal differential rotation determines the toroidal field generation, which has the highest dynamo efficiency around $\pm 55^\circ$ latitudes. The lower and higher latitudes have weaker latitudinal differential rotation, and hence a weaker dynamo efficiency. 
The latitudinal rotational shear and equatorward meridional flow contribute to the extended solar cycle from approximately $\pm 55^\circ$ latitudes to the equator, which lasts for about 18 years. The consecutive cycles persist for approximately 16 years, resulting in a 16-year period of the equatorward deceleration band of the torsional oscillations.
 
Band (II) corresponds to the equatorward acceleration band with the migration time period of about 16 years. It has two sub-bands. Band (IIa) follows the time-latitude evolution of the strong toroidal field, when and where  $B_{\phi}$ is large and $\partial B_{\phi}/\partial \theta$ is small. Hence the fourth term of Equation (\ref{eq:LorenzForce1}) dominates $F_{\phi}$, which satisfies 
\begin{equation}
F_{\phi}\approx\frac{1}{4\pi}\frac{\cot\theta B_{\theta}B_{\phi}}{r}.
	\label{eq:force_4th_term}
\end{equation}
As can be seen in Figure \ref{fig:ExplTwoBranches_ZJ22}(b), in this region $B_{\phi}$ has an opposite sign to $\partial B_{\phi}/\partial \theta$, and $B_{\theta}$ has the same sign as that in Band (I). Hence $F_{\phi}$ is in the rotation direction and leads to the acceleration band. Band (IIb) is near the equator and again has the large $\partial B_{\phi}/\partial \theta$ and small $B_{\phi}$ so that Equation (\ref{eq:force_2nd_term}) is satisfied. But $B_{\theta}$ has an opposite sign to that in Band (I). Hence $F_{\phi}$ in Band (IIb) is also in the rotation direction and leads to the acceleration band. Bands (IIa) and (IIb) cover the latitudes from about $\pm 55^\circ$ latitude to the equator for about 16 years.

The ZJ22 dynamo model reproduces the two poleward branches originating from the $\pm 55^\circ$ latitudes, which is the most distinguishing feature from the other two models. The two branches have the cycle period of about 8 years. 
Band (III) marked on Figure \ref{fig:ExplTwoBranches_ZJ22}(a) corresponds to the strong poleward acceleration of the torsional oscillations, and Band (IV) is the weak poleward deceleration band. At high-latitudes, the toroidal field is small and its derivative is large, so terms with $\partial B_{\phi}/\partial \theta$ and $\partial B_{\phi}/\partial r$ are larger than terms with $B_{\phi}$. Meanwhile, the coefficient of the fourth term, $\cot\theta$, is large at high latitudes. In such case, three terms of Equation (\ref{eq:LorenzForce1}) dominate $F_{\phi}$, which satisfies
\begin{equation}
	F_{\phi}\approx\frac{1}{4\pi}\left[  B_r\frac{\partial B_{\phi}}{\partial r}+\frac{B_{\theta}}{r}\frac{\partial B_{\phi}}{\partial \theta}+\frac{\cot\theta B_{\theta}B_{\phi}}{r}\right].
\label{eq:force_Br_term}
\end{equation}
As shown in Figure \ref{fig:ExplTwoBranches_ZJ22}(b), for the second term, most poleward $B_{\theta}$ and $\partial B_{\phi}/\partial \theta$ have the same sign and they change sign every 11 years, which corresponds to the acceleration band. Part poleward $\partial B_{\phi}/\partial \theta$ overlaps with opposite $B_{\theta}$, which reproduces the deceleration band. Figure \ref{fig:ExplTwoBranches_ZJ22}(c) shows that the phase relationship between $B_{\phi}$ and $B_{\theta}$ is similar to that between $B_{\theta}$ and $\partial B_{\phi}/\partial \theta$. Therefore, the deceleration band is narrower and weaker than the acceleration band in the polar regions.

%The poleward migration of $B_{\phi}$ results from the latitude dependence of the latitudinal rotational shear for the  $B_{\phi}$ generation, which peaks at about the $\pm 55^\circ$ latitudes \citep{Zhang2024}. 
%\textbf{The poloidal magnetic flux is first wound up at the peak of the latitudinal rotational shear, $\pm 55^\circ$ latitudes. This is where the toroidal field first generates. At the higher latitudes, the toroidal field generates later.}
%\textbf{The amplification of toroidal field occurs at mid-latitudes, and takes more time to complete at lower latitudes \citep{Babcock1961}.}
The poleward migration of $B_{\phi}$ results from the latitude dependence of the latitudinal rotational shear for the $B_{\phi}$ generation \citep{Cameron2023, Zhang2024}. The poloidal magnetic field is wound up and amplified by the latitudinal rotational shear to generate $B_{\phi}$. The shear peaks at about $\pm 55^\circ$ latitudes, where the strong $B_{\phi}$ first appears. At higher latitudes, the generation of $B_{\phi}$ takes longer time. The time lag in the generation of $B_{\phi}$ forms the poleward migration of $B_{\phi}$ as presented by Figure 4(a) of JZ22. The mechanism, which was initially proposed by \cite{Babcock1961} but ignored by the community, also has contributions to the equatorward migration.
Conversely, the poleward branch of the torsional oscillations originating from the $\pm 55^\circ$ latitudes provides an evidence that the toroidal field is generated in the bulk of the convection zone by the latitudinal rotational shear. The toroidal field is expected to present the `$\textless$' type on the time-latitude diagram with the turn-point at the $\pm 55^\circ$ latitudes, as shown on Figure \ref{fig:ExplTwoBranches_ZJ22}(c).

Figure \ref{fig:ExplTO_CNC_DC} shows the time-latitude diagram of the torsional oscillations at 0.8$R_\odot$ and the corresponding toroidal field from the CNC04 model (left panel) and the DC99 model (right panel). In the CNC04 model, the deep penetration of the meridional flow (0.61$R_\odot$) causes the equatorward toroidal field, generated mainly by the radial shear in the tachocline, from the polar regions to the equator \citep{Nandy2002}. Thus, there are only the equatorward branches of the torsional oscillations. The equatorward deceleration band occurs along the boundaries of two consecutive cycles based on Equation (\ref{eq:force_2nd_term}). In the left panel of Figure \ref{fig:ExplTO_CNC_DC}, there are also two equatorward acceleration bands at low latitudes. Their generation is similar to Band (II) of the ZJ22 model. At high latitudes, although there is a strong poleward $B_r$, the absence of the poleward $B_\phi$ above the $\pm 55^\circ$ latitudes still cannot generate the poleward band. The correlated poleward $B_r$ and equatorward $B_\phi$ generate the roughly stationary acceleration band at the poles. 

The right panel of Figure \ref{fig:ExplTO_CNC_DC} shows the toroidal field of the DC99 model at a depth of 0.8$R_\odot$. The toroidal field concentrates in two regions. 
One is around the polar regions with a center at about $\pm 70^\circ$ latitudes due to the strong radial shear of the tachocline at the polar regions. The other is between the $\pm 20^\circ$ latitudes due to the flux transport by the meridional flow. Being different from the simple poloidal field configuration given by the ZJ22 and CNC04 models, the poloidal field of the DC99 model has a complex configuration throughout all latitudes. The distribution of the poloidal and toroidal field results in the two branches of the torsional oscillations: a poleward acceleration band originating at about $\pm 70^\circ$ latitudes and an equatorward deceleration band slightly below it. The acceleration poleward bands are generated by the strong poleward branch of the poloidal and toroidal field originating from about $\pm 70^\circ$ latitudes. The toroidal field propagates towards equator, leading to the deceleration equatorward branch at high latitudes.
At lower latitudes, the poloidal field is much weaker. The toroidal field and its derivative is small at mid-latitudes but large below $\pm 20^\circ$ latitudes, corresponding to the equatorward branch originating from $\pm 20^\circ$ latitudes.

%%%%%%%%%%%%%%%%%%%%%%%%%%%%%%%%%%%%%%%%

%%%%%%%%%%%%%%%%%%%%%%%%%%%%%%%%%%%%%%%%%%
%\newpage
\section{Conclusions} \label{sec:conclusion}

In the paper, we have assumed that the Sun’s torsional oscillation is driven by the Lorentz force of the Sun’s cyclically varying magnetic field generated by the large-scale dynamo process. The torsional oscillation, specifically the zonal acceleration, has been used to discriminate between the three BL-type dynamo models, ZJ22, CNC04, and DC99. A comparison of the torsional oscillations from the three models presents significant difference in generating the observed properties of torsional oscillations. A subsequent analysis of the results provides insight into the configuration and relation between the poloidal and toroidal magnetic fields, and further into the solar dynamo process.

%The configuration and evolution of the toroidal and poloidal fields lead the three models to reproduce the various propagating branches of the torsional oscillations.
The first property of the torsional oscillations we used to discriminate between the three models is the poleward and equatorward branches originating from about $\pm 55^\circ$ latitudes. The two propagating branches of the observed torsional oscillations have regular migration time periods, which is related to the extended solar cycle \citep{McIntosh2014}. The ZJ22 model, in which the toroidal field is mainly generated in the bulk of the convection zone and the poloidal field has a global dipolar structure during cycle minimum, reproduces the poleward and equatorward branches of the torsional oscillations from about $\pm 55^\circ$ with the migration time periods consistent with the observed one. That is, the equatorward and poleward branches have migration time periods of 18 yr and 9 yr, respectively. The CNC04 model, in which the toroidal field lacks the poleward branch in the bulk of the convection, fails to reproduce the poleward branch of torsional oscillations. In the DC99 model, the poleward branch of torsional oscillations originates from $\pm 70^\circ$ latitudes due to the poloidal and toroidal field moving from there to the poles. Both CNC04 and DC99 models fail to reproduce migration time periods of the torsional oscillations.

The configuration of magnetic field in the convection zone is suggested to be the reason for the poleward and equatorward branches of torsional oscillations. \cite{Covas2000} also reproduced the poleward and equatorward branches in their model, where the toroidal field had the similar `$<$' type as referred in Section \ref{sec:twoBranches}. Thus two branches of the torsional oscillations might provide a constraint on the magnetic field, especially on the toroidal field in the convection zone. This constraint on toroidal field further might imply that the primary seat of the solar dynamo could be located in the bulk of the convection zone. This is consistent with the inference given by \cite{Mahajan2019}. \cite{Durney2000} also concluded that the contribution of generating toroidal field in the bulk of the convection zone is significant in reproducing the torsional oscillation. However, it contradicts the one given by \cite{Kosovichev2019}, who infer that the primary seat of the solar dynamo is located in a high-latitude zone of the tachocline based on their helioseismic analysis of torsional oscillations. 

The alternating acceleration and deceleration bands is the second property of the torsional oscillations we used to discriminate between the three models. The DC99 dynamo model is capable of reproducing the property of alternating acceleration and deceleration bands near the surface. The CNC model is also able to produce this property to a certain extent. This is because of the phase relationship between the surface $B_{r}$ and $B_{\phi}$. The phase difference is about $\pi/2$ in the DC99 model, and it is between $0$ and $\pi/2$ at high latitudes in the CNC04 model. The ZJ22 model fails to reproduce the alternating acceleration zone due to the phase difference about $\pi$. Combined with the boundary condition, we suggest that the torsional oscillation could provide a constraint on the phase relationship about $\pi/2$ between the near surface $B_{r}$ and $B_{\phi}$.

The amplitudes of the torsional oscillations in the convection zone represent the third property of the torsional oscillations that have been used to discriminate between the three models. Observations indicate that the amplitude of the zonal acceleration is of the order of $10^{-8}$ m s$^{-2}$ \citep{Kosovichev2019}. The ZJ22 dynamo model produces the torsional oscillations at a comparable level. The CNC04 and DC99 models produce the torsional oscillations by 3 and 6 orders of magnitude larger than the observed amplitude, respectively. This could lead to the third constraint on the magnetic field in the convection zone, namely that a toroidal component of several hundred Gauss may be sufficient to produce the observed torsional oscillations. 

In conclusion, our study indicates that the original idea proposed by \cite{Durney2000} is an effective approach to discriminate between dynamo models. The model discrimination process also provides insight into the magnetic field in the convection zone, which is currently lacking observational constraints. Therefore, we propose that the approach be applied when a new dynamo model is developed. \linebreak

\noindent  We acknowledge financial support from the National Natural Science Foundation of China (Nos. 12173005 and 1235000016) and the National Key R\&D Program of China No. 2022YFF0503800.

\bibliography{sample631}{}
\bibliographystyle{aasjournal}


%\iffalse 

%\fi
\end{document}